\documentstyle[12pt]{article}

\newif\ifprelim
\def\mode{b }  

\def\baselinestretch{1.2}
\def\unlock{\catcode`@=11 }
\def\lock{\catcode`@=12 }

\def\marginnote#1{}
\let\ju=\marginnote
\def\mybiblabel#1{#1\hfil}
\unlock
\def\@bibitem#1{\def\@blbl{#1}\item\if@filesw \immediate\write\@auxout
     {\string\bibcite{#1}{\the\value{\@listctr}}}\fi\ignorespaces}
\lock
\def\blackfonts{
                \font\blackboard=msbm10 scaled\magstep1
                \font\blackboards=msbm8
                \font\blackboardss=msbm6}

\def\bigmode{b }
\def\draftmode{d }
\def\bdraftmode{bd }
\ifx\mode\undefined\read-1 to\mode\fi
\def\bigpage{
        \textheight 22.5 cm
        \topmargin -.5 cm
        \textwidth 16cm
        \oddsidemargin 0 in
        \evensidemargin 0 in
}
\def\doublepage{
        \parskip 6 pt
        \parindent 2em
        \twocolumn
        
        \let\small\relax
        \let\sl\it
        \sloppy
        \voffset=-2.54truecm
        \hoffset=-1.54truecm
        \flushbottom
        \leftmargini 2em
        \leftmarginv .5em
        \leftmarginvi .5em
        \marginparwidth 48pt
        \marginparsep 10pt
        \setlength{\columnsep}{2truecm}
        \setlength{\textwidth}{25.4truecm}
        \setlength{\textheight}{17truecm}
        \oddsidemargin .18truein
        \evensidemargin .17truein
}
\def\draftpage{
        \textheight 19.86075cm
        \topmargin -.5 cm
        \textwidth 13.66667cm
        \oddsidemargin 0cm
        \evensidemargin 0cm
        \marginparwidth 96pt
        \marginparsep 10pt
}
\def\smallblack{
        \def\blackfonts{
                \font\blackboard=msbm10
                \font\blackboards=msbm7
                \font\blackboardss=msbm5
        }
}
\def\marginnotes{
        \def\draftmarginnote##1{\marginpar{\raggedright\scriptsize\tt ##1}}
        \let\marginnote=\draftmarginnote
        \let\ju=\marginnote
}
\unlock
\def\draftnotice{
        \def\@oddhead{\hfil \smash{\Large\em DRAFT} \hfil
                \em \today\quad\militarytime \hfil
                \smash{\Large\em DRAFT} \hfil}
        \let\@evenhead\@oddhead
        \def\ps@plain{\let\@mkboth\@gobbletwo
                \def\@oddfoot{\hfil \smash{\Large\em DRAFT} \hfil \thepage
                \hfil \smash{\Large\em DRAFT} \hfil}
                \let\@evenfoot\@oddfoot}
        \def\ps@empty{\let\@mkboth\@gobbletwo
                \def\@oddfoot{\hfil \smash{\Large\em DRAFT} \hfil}
                \let\@evenfoot\@oddfoot}
        \pagestyle{plain}
}
\def\eqnlabels{
        \def\draftlabel##1{{\@bsphack\if@filesw {\let\thepage\relax
           \xdef\@gtempa{\write\@auxout{\string
              \newlabel{##1}{{\@currentlabel}{\thepage}}}}}\@gtempa
           \if@nobreak \ifvmode\nobreak\fi\fi\fi\@esphack}
                \gdef\@eqnlabel{##1}}
        \def\@eqnlabel{}
        \def\@vacuum{}
        \let\label=\draftlabel
        \def\@eqnnum{(\theequation)\rlap{\kern\marginparsep\tt\@eqnlabel}%
                \global\let\@eqnlabel\@vacuum}
}
\def\draftbib{
        \def\mybiblabel##1{\llap{\scriptsize\tt \@blbl\ }##1\hfil}
        \def\@lbibitem[##1]##2{%
                \def\@blbl{##2}\item[\@biblabel{##1}\hfill]\if@filesw
                {\def\protect##1{\string ##1\space}\immediate
                \write\@auxout{\string\bibcite{##2}{##1}}}\fi\ignorespaces}
        \def\@bibitem##1{\def\@blbl{##1}\item\if@filesw
\immediate\write\@auxout
                {\string\bibcite{##1}{\the\value{\@listctr}}}\fi\ignorespaces}
        }
\lock

\ifx\mode\bigmode
        \ifprelim
                \draftnotice
        \else
        \fi
        \bigpage
\else\ifx\mode\draftmode
        \draftpage
        \draftnotice
        \marginnotes
        \eqnlabels
	\draftbib
\else\ifx\mode\bdraftmode
        \bigpage
        \draftnotice
        \marginnotes
        \eqnlabels
	\draftbib
\else                           
        \ifprelim
                \draftnotice
        \else
        \fi
        \unlock
        \input art10.sty
        \lock
        \def\baselinestretch{1.3}
        \smallblack
        \doublepage
\fi
\fi
\fi

\batchmode
        \newfont{\footbbbfont}{msbm10}
\errorstopmode

\newif\ifamsf\amsftrue
        \ifx\footbbbfont\nullfont
        \amsffalse
\fi

\ifamsf
        \blackfonts
        \newfam\black
        \textfont\black=\blackboard
        \scriptfont\black=\blackboards
        \scriptscriptfont\black=\blackboardss
        \def\ZZ{{\fam\black\relax Z}}

\else            
        \def\ZZ{{Z \n{10} Z}}

\fi

\newcount\hour
\newcount\minute
\newtoks\amorpm
\hour=\time\divide\hour by 60
\minute=\time{\multiply\hour by 60 \global\advance\minute by-\hour}
\edef\standardtime{{\ifnum\hour<12 \global\amorpm={am}%
        \else\global\amorpm={pm}\advance\hour by-12 \fi
        \ifnum\hour=0 \hour=12 \fi
        \number\hour:\ifnum\minute<10 0\fi\number\minute\the\amorpm}}
\edef\militarytime{\number\hour:\ifnum\minute<10 0\fi\number\minute}

\newif\ifepsfloaded
\newif\iffigureexists

\ifeof 1 \epsfloadedfalse \else \epsfloadedtrue \fi
\ifepsfloaded \input epsf \fi

\def\checkex#1 {\relax
    \openin 1 #1
    \ifeof 1 \figureexistsfalse
    \else \figureexiststrue
    \fi \closein 1 }

\def\cpsbox#1#2{
        \ifepsfloaded
                \checkex #2
                \iffigureexists
                        \immediate\write16{(#2)}
                        \setlength{\epsfxsize}{#1}
                        \centerline{\epsfbox{#2}}
                \else
                        \immediate\write16{(#2 NOT FOUND!)}
                        \vbox to 2in{\hbox to #1 {\hss} \vss}
                \fi
        \else
                \immediate\write16{(NOT inputting #2; no epsf.tex)}
                \vbox to 2in{\hbox to #1 {\hss} \vss}
        \fi}
\unlock
\def\@citex[#1]#2{%
\if@filesw \immediate \write \@auxout {\string \citation {#2}}\fi
\@tempcntb\m@ne \let\@h@ld\relax \def\@citea{}%
\@cite{%
  \@for \@citeb:=#2\do {%
    \@ifundefined {b@\@citeb}%
      {\@h@ld\@citea\@tempcntb\m@ne{\bf ?}%
      \@warning {Citation `\@citeb ' on page \thepage \space undefined}}%
      {\@tempcnta\@tempcntb \advance\@tempcnta\@ne%
      \@tempcntb\number\csname b@\@citeb \endcsname \relax%
      \ifnum\@tempcnta=\@tempcntb 
        \ifx\@h@ld\relax%
          \edef \@h@ld{\@citea\csname b@\@citeb\endcsname}%
        \else%
          \edef\@h@ld{\ifmmode{-}\else--\fi\csname b@\@citeb\endcsname}%
        \fi%
      \else
        \@h@ld\@citea\csname b@\@citeb \endcsname%
        \let\@h@ld\relax%
      \fi}%
    \def\@citea{,\penalty\@highpenalty\,}%
  }\@h@ld
}{#1}}

\def\@citeb#1#2{{[#1]\if@tempswa , #2\fi}}
\def\@citeu#1#2{{$^{#1}$\if@tempswa , #2\fi }}
\def\@citep#1#2{{#1\if@tempswa , #2\fi}}

\def\bcites{         
        \unlock
        \let\@cite=\@citeb
        \lock
}

\def\upcites{         
        \unlock
        \let\@cite=\@citeu
        \lock
}

\def\plaincites{      
        \unlock
        \let\@cite=\@citep
        \lock
}

\let\@cite=\@citeb              

\def\refname{References}        

\def\thebibliography#1{\section*{\refname\@mkboth
  {\uppercase{\refname}}{\uppercase{\refname}}}\list
  {\@biblabel{\arabic{enumiv}}}{\settowidth\labelwidth{\@biblabel{#1}}%
    \let\makelabel\mybiblabel\leftmargin\labelwidth
    \advance\leftmargin\labelsep
    \usecounter{enumiv}%
    \let\p@enumiv\@empty
    \def\theenumiv{\arabic{enumiv}}}%
    \def\newblock{\hskip .11em plus.33em minus.07em}%
    \sloppy\clubpenalty4000\widowpenalty4000
    \sfcode`\.=1000\relax}

\def\@noitemerr{\@warning{Something's wrong--perhaps a missing
\string\item}\@ehc}

\def\sections{\unlock
\def\theequation{\thesection.\arabic{equation}}
\@addtoreset{equation}{section}
\@addtoreset{footnote}{section}
\lock
}

\def\footnotesections{\unlock
\@addtoreset{footnote}{section}
\lock
}

\def\subsections{\unlock
\def\theequation{\thesubsection.\arabic{equation}}
\@addtoreset{equation}{subsection}
\@addtoreset{footnote}{subsection}
\lock
}

\lock

\let\apost\aposteriori

\def\ibid{{\it ibid.\/}}
\def\ie{\hbox{\it i.e.\/}}

\def\Kahler{K\"ahler}

\def\noj#1,#2,{{\bf #1} (19#2)\ }
\def\jou#1#2,#3,{{\em #1\/ }{\bf #2} (19#3)\ }
\def\ann#1,#2,{{\em Ann.\ Physics\/ }{\bf #1} (19#2)\ }
\def\annmath#1,#2,{{\em Ann.\ Math\/ }{\bf #1} (19#2)\ }
\def\cmp#1,#2,{{\em Comm.\ Math.\ Phys.\/ }{\bf #1} (19#2)\ }
\def\cq#1,#2,{{\em Class.\ Quantum Grav.\/ }{\bf #1} (19#2)\ }
\def\cqg#1,#2,{{\em Class.\ Quantum Grav.\/ }{\bf #1} (19#2)\ }
\def\ijmp#1,#2,{{\em Int.\ J.\ Mod.\ Phys.\/ }{\bf A#1} (19#2)\ }
\def\jmp#1,#2,{{\em J.\ Math.\ Phys.\/ }{\bf #1} (19#2)\ }
\def\grg#1,#2,{{\em Gen.\ Rel.\ Grav.\/ }{\bf #1} (19#2)\ }
\def\mpl#1,#2,{{\em Mod.\ Phys.\ Lett.\/ }{\bf A#1} (19#2)\ }
\def\nc#1,#2,{{\em Nuovo Cim.\/ }{\bf #1} (19#2)\ }
\def\np#1,#2,{{\em Nucl.\ Phys.\/ }{\bf B#1} (19#2)\ }
\def\pl#1,#2,{{\em Phys.\ Lett.\/ }{\bf #1B} (19#2)\ }
\def\pla#1,#2,{{\em Phys.\ Lett.\/ }{\bf #1A} (19#2)\ }
\def\pr#1,#2,{{\em Phys.\ Rev.\/ }{\bf #1} (19#2)\ }
\def\prd#1,#2,{{\em Phys.\ Rev.\/ }{\bf D#1} (19#2)\ }
\def\prl#1,#2,{{\em Phys.\ Rev.\ Lett.\/ }{\bf #1} (19#2)\ }
\def\prp#1,#2,{{\em Phys.\ Rept.\/ }{\bf #1C} (19#2)\ }
\def\ptp#1,#2,{{\em Prog.\ Theor.\ Phys.\/ }{\bf #1} (19#2)\ }
\def\ptpsup#1,#2,{{\em Prog.\ Theor.\ Phys.\/ Suppl.\/ }{\bf #1} (19#2)\ }
\def\rmp#1,#2,{{\em Rev.\ Mod.\ Phys.\/ }{\bf #1} (19#2)\ }
\def\yadfiz#1,#2,#3[#4,#5]{{\em Yad.\ Fiz.\/ }{\bf #1} (19#2) #3%
\ [{\em Sov.\ J.\ Nucl.\ Phys.\/ }{\bf #4} (19#2) #5]}
\def\zh#1,#2,#3[#4,#5]{{\em Zh.\ Exp.\ Theor.\ Fiz.\/ }{\bf #1} (19#2) #3%
\ [{\em Sov.\ Phys.\ JETP\/ }{\bf #4} (19#2) #5]}

\def\eq#1{.~(\ref{#1})}
\def\noeq#1{(\ref{#1})}
\def\beq{\begin{equation}}
\def\eeq{\end{equation}}
\def\beqar{\begin{eqnarray}}
\def\eeqar{\end{eqnarray}}

\def\beqal{\begin{equation}\begin{eqalign}}
\def\eeqal{\end{eqalign}\end{equation}}
\def\beqaltwo{\begin{eqaligntwo}}
\def\eeqaltwo{\end{eqaligntwo}}

\def\p#1{\mskip#1mu}
\def\n#1{\mskip-#1mu}
\def\stop{\p6.}
\def\comma{\p6,}
\def\semi{\p6;}

\def\excl{\p6!}

\def\to{\rightarrow}

\def\longlongrightarrow{\relbar\joinrel\relbar\joinrel\rightarrow}

\def\onArrow#1{\mathrel{\mathop{\longlongrightarrow}\limits^{#1}}}

\def\nfrac#1#2{{\displaystyle{\vphantom1\smash{\lower.5ex\hbox{\small$#1$}}%
        \over\vphantom1\smash{\raise.25ex\hbox{\small$#2$}}}}}

\def\pa{\partial}
\def\pb{\bar\pa}

\def\Tr{{\rm Tr}}
\def\l:{\mathopen{:}\,}
\def\r:{\,\mathclose{:}}

\unlock
\def\@versim#1#2{\smash{\lower0.5ex\vbox{\baselineskip\z@skip\lineskip\z@skip
        \lineskiplimit\z@\ialign{$\m@th#1\hfil##\hfil$\crcr#2\crcr\sim\crcr}}}}
\def\ltsim{\mathrel{\mathpalette\@versim<}}
\def\gtsim{\mathrel{\mathpalette\@versim>}}
\lock

\def\cald{{\cal D}}

\def\calf{{\cal F}}

\def\calm{{\cal M}}

\def\cals{{\cal S}}

\def\abar{{\bar a}}

\def\ibar{{\bar i}}

\def\qbar{{\bar q}}

\def\tbar{{\bar t}}

\def\zbar{{\bar z}}
\def\Abar{{\bar A}}

\def\Pbar{{\bar P}}
\def\Qbar{{\bar Q}}

\def\Xbar{{\bar X}}
\def\Ybar{{\bar Y}}

\def\Ttilde{{\tilde T}}

\def\mubar{{\bar \mu}}
\def\nubar{{\bar \nu}}

\def\rhobar{{\bar \rho}}
\def\sigmabar{{\bar \sigma}}

\def\rhotilde{{\tilde \rho}}

\unlock
\def\section{\@startsection {section}{1}{\z@}{3.ex plus 1ex minus
 .2ex}{2.ex plus .2ex}{\large\bf}}
\def\subsection{\@startsection{subsection}{2}{\z@}{2.75ex plus 1ex minus
 .2ex}{1.5ex plus .2ex}{\bf}}
\if@twocolumn
                \def\abstractspace{\vskip .5 cm}
                \def\abstractindent{1.5 cm}
\else
                \def\abstractspace{}
                \def\abstractindent{.75 cm}
\fi

\def\appendix{{\newpage\section*{Appendices}}\let\appendix\section%
        {\setcounter{section}{0}
        \gdef\thesection{\Alph{section}}}\section}

\def\abstract{\abstractspace\begin{center}
        {\bf Abstract}
        \end{center}
        \advance\leftskip\abstractindent
        \advance\rightskip\abstractindent
}

\lock

\unlock

\newif\if@defeqnsw \@defeqnswtrue

\def\eqnarray{\stepcounter{equation}\let\@currentlabel=\theequation
\if@defeqnsw\global\@eqnswtrue\else\global\@eqnswfalse\fi
\global\@eqnswtrue
\tabskip\@centering\let\\=\@eqncr
$$\halign to \displaywidth\bgroup\hfil\global\@eqcnt\z@
  $\displaystyle\tabskip\z@{##}$&\global\@eqcnt\@ne
  \hfil$\displaystyle{{}##{}}$\hfil
  &\global\@eqcnt\tw@ $\displaystyle{##}$\hfil
  \tabskip\@centering&\llap{##}\tabskip\z@\cr}

\def\yesnumber{\global\@eqnswtrue}

\def\@@eqncr{\let\@tempa\relax\global\advance\@eqcnt by \@ne
    \ifcase\@eqcnt \def\@tempa{& & & &}\or \def\@tempa{& & &}\or
     \def\@tempa{& &}\or \def\@tempa{&}\else\fi
     \@tempa \if@eqnsw\@eqnnum\stepcounter{equation}\fi
     \if@defeqnsw\global\@eqnswtrue\else\global\@eqnswfalse\fi
     \global\@eqcnt\z@\cr}

\def\@eqnacr{{\ifnum0=`}\fi\@ifstar{\@yeqnacr}{\@yeqnacr}}

\def\@yeqnacr{\@ifnextchar [{\@xeqnacr}{\@xeqnacr[\z@]}}

\def\@xeqnacr[#1]{\ifnum0=`{\fi}\cr \noalign{\vskip\jot\vskip #1\relax}}

\def\eqalign{\null\,\vcenter\bgroup\openup1\jot \m@th \let\\=\@eqnacr
\ialign\bgroup\strut
\hfil$\displaystyle{##}$&$\displaystyle{{}##}$\hfil\crcr}
\def\endeqalign{\crcr\egroup\egroup\,}

\def\cases{\left\{\,\vcenter\bgroup\normalbaselines\m@th \let\\=\@eqnacr
    \ialign\bgroup$##\hfil$&\quad##\hfil\crcr}
\def\endcases{\crcr\egroup\egroup\right.}

\def\eqalignno{\stepcounter{equation}\let\@currentlabel=\theequation
\if@defeqnsw\global\@eqnswtrue\else\global\@eqnswfalse\fi
\let\\=\@eqncr
$$\displ@y \tabskip\@centering \halign to \displaywidth\bgroup
  \global\@eqcnt\@ne\hfil
  $\@lign\displaystyle{##}$\tabskip\z@skip&\global\@eqcnt\tw@
  $\@lign\displaystyle{{}##}$\hfil\tabskip\@centering&
  \llap{\@lign##}\tabskip\z@skip\crcr}

\def\endeqalignno{\@@eqncr\egroup
      \global\advance\c@equation\m@ne$$\global\@ignoretrue}

\@namedef{eqalignno*}{\@defeqnswfalse\eqalignno}
\@namedef{endeqalignno*}{\endeqalignno}

\def\eqaligntwo{\stepcounter{equation}\let\@currentlabel=\theequation
\if@defeqnsw\global\@eqnswtrue\else\global\@eqnswfalse\fi
\let\\=\@eqncr
$$\displ@y \tabskip\@centering \halign to \displaywidth\bgroup
  \global\@eqcnt\m@ne\hfil
  $\@lign\displaystyle{##}$\tabskip\z@skip&\global\@eqcnt\z@
  $\@lign\displaystyle{{}##}$\hfil\qquad&\global\@eqcnt\@ne
  \hfil$\@lign\displaystyle{##}$&\global\@eqcnt\tw@
  $\@lign\displaystyle{{}##}$\hfil\tabskip\@centering&
  \llap{\@lign##}\tabskip\z@skip\crcr}

\def\endeqaligntwo{\@@eqncr\egroup
      \global\advance\c@equation\m@ne$$\global\@ignoretrue}

\@namedef{eqaligntwo*}{\@defeqnswfalse\eqaligntwo}
\@namedef{endeqaligntwo*}{\endeqaligntwo}

\newtoks\@stequation

\def\subequations{\refstepcounter{equation}%
  \edef\@savedequation{\the\c@equation}%
  \@stequation=\expandafter{\theequation}
  \edef\@savedtheequation{\the\@stequation}
  \edef\oldtheequation{\theequation}%
  \setcounter{equation}{0}%
  \def\theequation{\oldtheequation\alph{equation}}}

\def\endsubequations{%
  \setcounter{equation}{\@savedequation}%
  \@stequation=\expandafter{\@savedtheequation}%
  \edef\theequation{\the\@stequation}%
  \global\@ignoretrue}

\def\big#1{{\hbox{$\left#1\vcenter to1.428\ht\strutbox{}\right.\n@space$}}}
\def\Big#1{{\hbox{$\left#1\vcenter to2.142\ht\strutbox{}\right.\n@space$}}}
\def\bigg#1{{\hbox{$\left#1\vcenter to2.857\ht\strutbox{}\right.\n@space$}}}
\def\Bigg#1{{\hbox{$\left#1\vcenter to3.571\ht\strutbox{}\right.\n@space$}}}

\lock

\def\CY{Calabi--Yau}
\def\CS{Chern--Simons}
\def\KS{Kodaira--Spencer}
\def\RS{Ray--Singer}
\def\einbein{{\it einbein\/}}
\def\einbeins{{\it einbeins\/}}

\def\utwo{$U(2)$}
\def\uone{$U(1)$}
\def\dtwoz{{\rm d}^{\p1 2}\n2}
\def\dT{{\rm d}T}
\def\dTtilde{{\rm d}\Ttilde}
\def\dt{{\rm d}t}
\def\dTT{\frac\dT T}

\def\dTTs{\frac\dT {T^{\, 1-s}}}
\def\dTTTs{\frac\dT {T^{\, 2-s}}}
\def\Yi{Y_i}
\def\Ybi{\Ybar_\ibar}
\def\sump{\smash{{\sum_{n,m}} ^{\;\prime}}}
\def\sumnm{\sum_{n,m}}
\def\Q#1{Q^{\p2 #1}}
\def\Qb#1{\Qbar_{\p2 #1}}

\def\e{{\rm e}}
\def\diag{{\rm diag}}
\def\det{{\rm det}}
\def\sdet{{\rm sdet}}

\footnotesections

\begin{document}
\begin{titlepage}

\noindent August 11, 1994\hfill CERN--TH.7402/94\\
\null\hfill TAUP--2192--94\\
\null\hfill hep-th/9408116

\vskip .5 cm

\begin{center}

{\large \bf
The topological B model as a twisted spinning particle}

\vskip .3 cm
{
Neil Marcus$^{(a)\,*}$ {\em and\/}
Shimon Yankielowicz$^{(a,b)\,}$\footnote{
Work supported in part by the US-Israel Binational Science Foundation,
the German-Israeli Foundation for Scientific Research and Development
and the Israel Academy of Science.  E-Mail:
NEIL@HALO.TAU.AC.IL, H75@TAUNIVM.TAU.AC.IL}
}

\vskip 0.15 cm

${}^{(a)}${\sl%
School of Physics and Astronomy\\Raymond and Beverly Sackler Faculty
of Exact Sciences\\Tel-Aviv University\\Ramat Aviv, Tel-Aviv 69978, ISRAEL.
}
\vskip 0.15 cm

${}^{(b)}${\sl%
Theory Division, Cern\\
               CH--1211 Geneva 23, Switzerland
}

\end{center}

\begin{abstract}
\vskip -0.15 cm

The B--twisted topological sigma model coupled to topological gravity
is supposed to be described by an ordinary field theory: a type of
holomorphic \CS{} theory for the open string, and the \KS{} theory for
the closed string.  We show that the B model can be represented as a
{\em particle\/} theory, obtained by reducing the sigma model to one
dimension, and replacing the coupling to topological gravity by a
coupling to a twisted one-dimensional supergravity.  The particle can
be defined on {\em any\/} \Kahler{} manifold---it does not require the
\CY{} condition---so it may provide a more generalized setting for the
B model than the topological sigma model.

The one-loop partition function of the particle can be written in terms
of the Ray--Singer torsion of the manifold, and agrees with that of the
original B model.  After showing how to deform the \Kahler{} and
complex structures in the particle, we prove the independence of this
partition function on the \Kahler{} structure,
and investigate the origin of the holomorphic anomaly.  To define other
amplitudes, one needs to introduce interactions into the particle.  The
particle will then define a field theory, which may or may not be the
\CS{} or \KS{} theories.

\end{abstract}
\newpage

\end{titlepage}

\section{Introduction}

Witten showed that by twisting an $N=(2,2)$ sigma model one obtains a
topological theory.  In fact, depending on  the relative sign of the $U(1)$
charges used to twist the theory, one obtains two topological theories: the
A and the B models.  From a world-sheet point of view the two theories are
very similar, and ``mirror symmetry'' relates the
A model on one manifold to the B model on its mirror \cite{B}.
This has proven
to be very useful for \CY{} calculations,
since the basic observables of A
models correspond to \Kahler{} deformations of the manifold \cite{A}, while
those of the B models correspond to complex-structure deformations.

Despite this apparent similarity of the A and B
model, they have many basic differences.
One such difference is that the B twisting is
chiral, so the theory has a world-sheet Lorentz anomaly if the target space
is not a \CY{} manifold, which has a vanishing first Chern class, $c_1$
\cite{B}.  The A model, on the other hand, can be defined on any \Kahler{}
space\footnote{They can actually be defined on any almost-complex space,
although one only has two BRST symmetries in the \Kahler{} case, and only there
is the model a twisting of a $(2,2)$ sigma model \cite{A}.}, with the
restriction to being \CY{} necessary only for conformal
invariance.  Thus the \CY{} condition seems far more basic in the B model, and
mirror symmetry can apparently only exist in the (physically relevant) \CY{}
case.

A more surprising difference is that the target space interpretations of the
two theories seem to be completely different.  The bosonic part of both
theories, coming from the original $(2,2)$ sigma model, has the form
\beq
\cals =\int \dtwoz z \left( \,  t^a \, k_{\mu\mubar}^{(a)} \, \partial_z X^\mu
\partial_\zbar \Xbar^\mubar + \tbar^\abar \, k_{\mu\mubar}^{(a)} \,
\partial_\zbar X^\mu \partial_z \Xbar^\mubar \, \right) \comma \label{sigma}
\eeq
where $X^\mu$ is a complex coordinate on the target space, and the
$k_{\mu\mubar}^{(a)}$'s are a normalized basis of \Kahler{} metrics. The
$t^a$'s are coordinates on the moduli space of the complexified \Kahler{}
deformations of the theory; thus the target-space metric is
$G_{\mu\mubar}= Re (t^a) \, k_{\mu\mubar}^{(a)}$, and the  antisymmetric
target-space tensor on the space is
$B_{\mu\mubar}= Im (t^a) \, k_{\mu\mubar}^{(a)}$.  In Witten's
original study of the A model he showed that the (supersymmetrized)
$\tbar$ term is BRST
exact \cite{A}.  He then argued that one could study the theory in the
formal limit $\tbar \to \infty$, which enforces the condition that
maps from the worldsheet to the target space must be holomorphic.  The
$t^a$ term then reduces to the instanton number of the map, and the A model
simply ``counts'' holomorphic maps or, more generally, evaluates the Euler
characteristic of the moduli space of such maps.  Recently  Bershadsky,
Cecotti, Ooguri and Vafa (BCOV) discovered that when the A model is coupled to
topological gravity---which is necessary if one wants interesting loop
amplitudes---there is a BRST anomaly in the theory, so that it actually
does
depend on $\tbar$\footnote{See also ref.~\cite{kapl}, where this
is related to the non-holomorphicity of threshold corrections in the
string.} \cite{F1}.  However, one can still study the ``traditional'' A
model with ``base-point'' $\tbar \to \infty$, where the theory does
have a topological target-space interpretation.

The B model is far less well understood.  In it, \Kahler{} deformations are
BRST exact \cite{B}, so one can study it in the limit where both $t$ and
$\tbar$ become infinite. (More physically, one takes the large volume limit,
where $V \propto t+\tbar\to \infty$.)  It thus appears that the B model is
concerned only with constant maps, and evaluates only quantities depending on
the classical geometry of the target space.  However, the situation is again
complicated by the coupling of the theory to topological gravity.  In this case
BCOV showed that there is essentially no anomalous dependence on $t$ or
$\tbar$.  However, the coupling means that one must integrate amplitudes
over the moduli space of world-sheet Riemann surfaces.  As was noticed by
Witten, this means that the suppression of the action due to the large
target-space volume $V$ can be counteracted by being near to a degeneration
of the Riemann surface \cite{CS}.  This is most easily seen in the
hamiltonian quantization of the theory, which is relevant for the one-loop
amplitude.  Thus, consider a world-sheet torus, with a coordinate $\sigma$
ranging from 0 to 1 along the string, and a coordinate $t$ (not to be
confused with the $t^{(a)}$'s!) ranging from 0 to $T$, measuring the
proper time along the worldsheet.  In the large volume limit $V \to
\infty$ the antisymmetric tensor $B_{\mu\mubar}$ becomes irrelevant, and
the action \noeq{sigma} reduces to
\beq
\cals \to \int \dt
\, d \sigma \; G_{\mu\mubar} \left(\, \dot \Xbar{}^\mubar \, \dot X^\mu +
\Xbar^{\prime\,\mubar} \, X^{\prime\,\mu}  \, \right) \stop
\eeq
The first term is proportional to $V/T$, whereas the second goes like $V\, T$.
Thus in the limit $V \to \infty$, there is a contribution to the path integral
when $T\to \infty$ with $T/V$ constant.  Such a torus is conformally
equivalent to a circle---the worldline of a particle, instead of a
string---and one sees that in this limit the amplitude is dominated by
particle-like
excitations.  The straightforward generalization of this argument to arbitrary
genus amplitudes shows that the partition function of the B model, which can be
calculated in the limit $V \to \infty$, is dominated by configurations where
the string worldsheet collapses to a Feynman-diagram-like structure, so it is
natural to assume that B models should be calculable as (relatively) ordinary
field theories.

The rest of this paper is organized as follows:  In the next section we
present the $N=(2,2)$ sigma model reduced to one dimension and  its
symmetries.  In section 3 we discuss how to couple the theory to gravity,
to obtain a ``spinning particle''.  In section 4 we perform the path
integral of this particle on the circle, in the case when the target space
is a complex one torus.
(This example will be considered repeatedly throughout the
paper, being  both tractable and very instructive.)
We connect the appearance of the holomorphic anomaly in this case to a
conflict with modular invariance.
Using the torus result, we derive the partition function on an arbitrary
\Kahler{} manifold in section 5 using hamiltonian quantization, and relate
it to the \RS{} torsion on the manifold.  We find that the Hilbert space
of the particle is described in terms of $(p,q)$ forms, which is
different from the $(0,q)$ forms in $\wedge^p \, T(M)$ in Witten's
cohomology calculations, except on \CY{} manifolds.
In section 6, we discuss how to
vary the \Kahler{} and complex structures of the manifold in the particle
case.
We derive the \KS{} equation, and also find an apparently necessary
auxiliary condition that complex-structure variations must satisfy!
In section 7, we show the independence of the particle on the \Kahler{}
structure, and in section 8 we investigate
the holomorphic anomaly.  Finally, we present our conclusions, and close with
some speculations on the further development of the particle and its associated
field theory.

\section{A particle theory for the B model}

We have given the argument as to why the B--string should be describable as a
field theory: the next step is to find it!
There is no completely deductive procedure for constructing a string field
theory from a two-dimensional description of
a string.  Generally, the only clue is that the equation of
motion of the field theory should correspond to the allowed states and
deformations of the string.  For the
open string, Witten argued that the appropriate field theory should have the
general structure of a \CS{} theory \cite{CS}.  This was partially based on the
structure of open string field theory \cite{OS}, and partially on knowing
that to preserve the BRST symmetry, one can couple the string
only to connections with vanishing $(0,2)$ curvature \cite{CS}.
For the closed string, BCOV constructed a ``\KS'' field theory \cite{KS}.
This was based on the fact that the observables in the B model are the
deformations of the complex structure \cite{B}, so the string equation of
motion should give the \KS{} equation \cite{\KS}, which describes such
deformations.  The purpose of our work is to give a relatively deductive
derivation of the particle---as opposed to field theory---interpretation of
the B--string.  Such a particle theory is directly capable of giving only
the propagators and partition function of the theory, and interactions will
later need to be incorporated in order to reproduce the Feynman diagrams of
the field theory and the string.

Temporarily setting aside the coupling to topological gravity, the particle
action can be derived by taking the string action of the B model on the torus,
and dimensionally reducing the theory to a circle.  (Of course, at this point,
one needs the full two-dimensional action \cite{B},
including all the fermionic
terms.)  However, in practice it is better to construct the particle
action directly, rather than by dimensional reduction.  This is because the
notion of spin becomes irrelevant in one dimension, so the fermionic fields
become scalars.  Thus the dimensional reductions of the $A$
\cite{A} or B \cite{B} models are the same as that of the untwisted $(2,2)$
sigma model \cite{22}, and since the distinction between left- and right-moving
fields is also lost in one dimension, the $U(1)_L \otimes U(1)_R$ symmetry
of the
sigma model is enhanced to a \utwo{} symmetry in the particle.  Knowing this,
one is lead uniquely to the usual so-called $N=1$ one-dimensional sigma model.
This sigma model can be written on any
Riemann manifold, where it has an $O(2)$ symmetry.
When the target space is \Kahler, the symmetry is enhanced to a
\utwo{}.  Introducing a dimensionful
coupling $\hbar$, its action can be written\footnote{The
transcription from the B model of ref.~\cite{B} to our notation is
$\rho_t \to \chi_1$, $\rho_x \to \chi_2$, $\eta \to \chi^{* \, 1}$ and
$\theta \to \chi^{* \, 2}$.  From section \ref{Hamsec} on we shall return
to the notation of \cite{B}, except for replacing $\rho_t \to \rho$ and
$\rho_x \to \rhotilde$.}:
\beq
\cals =  \frac1\hbar
        \int \dt \; G_{\mu\mubar} \,\dot \Xbar{}^\mubar \,  \dot X ^\mu
        + i \, \chi_\mu^{*\,i} \, \dot \chi_i^\mu + i \, \chi_\mu^{*\,i} \,
\Gamma_{\rho\sigma}^\mu \dot X ^\rho \chi_i^\sigma - \, \nfrac12 \,
R_{\mu\mubar\nu\nubar} \, \chi_j^\mu \, \chi^{* \, j \mubar} \, \chi_i^\nu \,
\chi^{*\,i\nubar} \comma \label{nograv}
\eeq
with the \utwo{} acting manifestly on the $i$ and $j$ indices of
the fermionic fields.  In \Kahler{} space the action is invariant under two
complex global supersymmetries, with parameters $\alpha_i$.  Formally taking
$\alpha_i$ and its complex
conjugate $\alpha^{*\,i}$ to be independent, the $\alpha_i$ transformations are
given by:
\beqaltwo
&\delta \, \Xbar^\mubar = -i \, \alpha_i \,  \chi^{* \, i \mubar}
&&\delta \, X^\mu =0 \nonumber \\
&\delta \, \chi_i^\mu = \dot X^\mu \, \alpha_i
&&\delta \, \chi_\mu^{*\,i} = 0  \comma \label{super}
\eeqaltwo
and the $\alpha^{*\,i}$ transformations by the complex conjugate
of \noeq{super}\footnote{The action is hermitian, up to
integrations by parts, and so should be completely symmetrical with respect
to the ``starred'' and ``unstarred'' fields.  Our choice to lower the
space-time indices on the spinors
$\chi^*$'s obscures this symmetry.  In particular, the $\alpha^{*\,i}$
transformations on the spinors is somewhat complicated (see eq\eq{G'}).}.
This action is
indeed the particle version of the B model closed string, before
coupling to topological gravity.  In particular, its partition
function $\Tr (-1)^F$ is simply the Euler number of the target space.

In the case of the open string, one sees that $\chi_2$ and $\chi^{* \, 2}$
have
antiperiodic boundary conditions, so they must be dropped in the particle
limit.  The particle version of the open B--string \cite{CS} is therefore
simply the $U(1)$ truncation of \noeq{nograv}.  This should describe a type
of \CS{} field theory, whose solutions are connections $A$ in the Chan--Paton
gauge group with vanishing $(0,2)$ curvature \cite{CS}.  To write the theory
with these background fields included, one needs to introduce Wilson lines of
the improved pullback of the connection
$\Phi^*(A) -i \, \eta^\mubar \, F_{\mubar \nu} \, \rho^\nu$
into the path-integral of the string or the particle
\cite{CS}. If one rewrites these Wilson lines in terms of
an integral over ``boundary fermions'' \cite{boundfer},
one remains with a description of
the theory in terms of a particle action.

\section{Coupling to gravity}

So far we have given simple arguments leading to a unique action for the matter
B model.  However, there is no deductive procedure for coupling the theory to
topological gravity.  In fact, even the original B sigma-model action has not
been explicitly coupled to topological gravity.  The
only indication of this coupling is given by the form of the amplitudes of the
theory.  BCOV argued, by analogy to the form of ghost insertions in the
bosonic string, that the one-loop partition function of the theory coupled to
topological gravity should be given by \cite{F1}:
\beq
\calf_1= \frac12 \, \int_\calm {d^2\tau \over \tau_2}\;
\Tr \, (-1)^F F_L  \,F_R  \; q^{H_L} \,
\qbar^{H_R} \comma
\label{F1}
\eeq
which is their ``generalized index'' \cite{index,ising}\footnote{We should
perhaps note here that $\calf_1$ is infinite, because of the
zero-modes of the hamiltonian.  Only differences or derivatives of
$\calf_1$ are well defined.}.  The integration over $\tau$ comes, as
usual, from writing the model on an arbitrary curved worldsheet, and
integrating over the metric, modulo diffeomorphisms and Weyl transformations.
The analogous coupling of
the particle to the one-dimensional \einbein{} $e$ can be found either by gauge
fixing the two-dimensional metric to $g_{\alpha\beta} \to \diag \left( e^2 ,1
\right)$, or by inserting \einbeins{} to make the action invariant under
one-dimensional diffeomorphisms.  Note that we do not have the Liouville
modes that are crucial in two-dimensional topological
gravity \cite{A,topgrav,DVV}.  This means \cite{DVV} that we will not find
gravitational descendants in our approach.

At this stage,
the only effect of the other fields of the
topological gravity is to give the insertions of  $F_L$ and $F_R$ into the
partition function.  It is natural, but incorrect, to attempt to introduce
these other fields by gauging all the global symmetries of the
action, \ie{} the \utwo{} symmetry and the four supersymmetries.  (This
does give a new type of \Kahler{} spinning particle theory, which is
interesting in its own right \cite{to appear}.)  To find the correct
procedure, it is useful  to track the appearance of these symmetries from the
original B model, in
order to decide which of them to gauge.  As we have stated before, the diagonal
subgroup of the \utwo{} comes from the $U(1)_L$ and $U(1)_R$ in the original
theory; the rest of the group is not a symmetry of the string.  The two
supersymmetries of eq\eq{super} come from the left- and right-handed BRST
symmetries of the B model, while the complex-conjugate supersymmetries come
from the sigma-model symmetries generated by the $G_{zz}$ and $G_{\zbar\zbar}$
of the two $N=2$ superconformal algebras.  (Since spin no longer
has any meaning, there
is no distinction between a supersymmetry and a BRST invariance in the
particle.)  In the original sigma model one would not gauge the BRST
symmetries or the \uone's, so we shall gauge only diffeomorphisms and the
two ``$G$'' symmetries\footnote{A similar coupling of the B sigma
model to topological gravity was considered in \cite{lamba}.}.  This
asymmetrical choice means that the theory is no longer unitary, and that we
have finally distinguished the B model from the untwisted $(2,2)$ sigma
model!

Proceeding to gauge the  $\alpha^{*\,i}$ symmetries with
gravitini $\psi^{*\,i}$, we obtain our action
\beqal
\cals =    \frac1\hbar
        \int \dt \; &
        \frac1e \; G_{\mu\mubar} \, \dot \Xbar{}^\mubar
              \left( \, \dot X ^\mu + i \, \psi^{*\,i} \; \chi_i^\mu \, \right)
        + i \, \chi_\mu^{*\,i} \dot \chi_i^\mu + i \, \chi_\mu^{*\,i} \,
        \Gamma_{\rho\sigma}^\mu \dot X ^\rho \chi_i^\sigma \\
        & - \, \nfrac{e}2\,
        R_{\mu\mubar\nu\nubar} \, \chi_j^\mu \, \chi^{* \, j \mubar} \,
                                        \chi_i^\nu \, \chi^{*\,i\nubar}
\stop \label{particle}
\eeqal
Note that, unlike the ungauged theory, $\cals$ can be written only
on \Kahler{} manifolds.
However, since particle theories never have local anomalies, being field
theories in
an odd number of dimensions, {\em there is no reason to impose the
vanishing of $c_1$, so one is not restricted to \CY{} manifolds.\/}

The action $\cals$
is clearly diffeomorphism invariant, and has a manifest global
\utwo{}.  In the gauged theory, the
two supersymmetries of \noeq{super} become
\beqaltwo
&\delta \, e =  -i \, \alpha_i \, \psi^{*\,i}
&&\delta \, \psi^{*\,i} = 0 \nonumber \\
&\delta \, \Xbar^\mubar = -i \, \alpha_i \, \chi^{* \, i \mubar}
&&\delta \, X^\mu = 0   \label{BRST} \\
&\delta \, \chi_i^\mu = \nfrac1e \left( \, \dot X^\mu + i \, \psi^{*\,j} \,
        \chi_i^\mu \, \right) \alpha_i
&&\delta \, \chi_\mu^{*\,i} = 0 \nonumber \stop
\eeqaltwo
These can be recognized as the
remnants of the BRST transformations of the B model
\cite{B}, and of topological gravity \cite{DVV}.
The two local supersymmetries with some spinor indices raised or lowered
are given by
\beqaltwo
&\delta \, \psi^{*\,i} = \dot \alpha^{*\,i}
&\qquad& \delta  \, e = 0 \nonumber \\
&\delta \, X^\mu = -i \, \alpha^{*\,i} \, \chi_i^\mu
&&\delta \, \Xbar^\mubar = 0 \label{G} \\
&\delta \, \chi^{*\,i\mubar} = \nfrac1e \, \dot \Xbar{}^\mubar \, \alpha^{*\,i}
&&\delta \, \chi_{i\mubar}  = 0 \nonumber \stop
\eeqaltwo
The transformations of the spinors with the original indices
are somewhat more complicated:
\beqaltwo
&\delta \, \chi_\mu^{*\,i} = \nfrac1e \, G_{\mu\mubar} \,
        \dot \Xbar{}^\mubar \, \alpha^{*\,i} - i \, \Gamma_{\mu\sigma}^\rho \,
        \alpha^{*\,j} \chi_j^\sigma  \chi_\rho^{*\,i}
&&\delta \, \chi_i^\mu = i \, \Gamma_{\rho\sigma}^\mu \,
        \alpha^{*\,j} \chi_j^\sigma  \chi_i^\rho \comma \label{G'}
\eeqaltwo
and include the noncovariant looking terms involving
the Christoffel symbols that usually appear in spinor
transformations in supersymmetric sigma models.

To get the ``open string'' particle, one
should truncate eqs.~(\ref{particle}--\ref{G'}) to the $U(1)$ case.

\section{The particle on a complex torus}

Having written a particle action for the B model, we would now like to
evaluate its partition function.  As usual, the diffeomorphisms
on the circle can be gauge-fixed by setting $\dot e = 0$, up to constant
translations and the $\ZZ_2$ symmetry of inverting the circle.  Thus the
path integral over the metric reduces to an integral over the length $T$ of
the circle, with measure \cite{polybook}
\beq
\frac12 \, \int_0^\infty \dTT \, T^s \stop \label{einb}
\eeq
Here we have introduced a proper time regulator $T^s$ for the
ultraviolet and infrared infinities of the theory.  In this
zeta-function-like  regularization one first writes all quantities
as meromorphic functions in $s$, and then lets $s \to 0$ discarding all poles.

One would similarly like to choose the gauge $\dot \psi{}^{*\,i} = 0$ for the
gravitini.
Unfortunately here one runs into a difficulty: As with the diffeomorphisms, the
gauge does not fix the local
supersymmetry transformations \noeq{G} with constant $\alpha^{*\,i}$'s.
These are analogues of conformal Killing spinors in fermionic
strings, and they are difficult to deal with, since one
must divide the path integral by the volume of the transformation group,
which is zero.  Usually one is saved from having to
perform the calculation by an overcompensation of fermion-matter zero modes
in the numerator of the integral, and the only previous case in which such a
calculation was really needed is in the one-loop amplitude of the $N=2$ string
\cite{MM}.  In that string one has to integrate over \uone{} moduli, so the
calculation could be done by considering the theory with twisted boundary
conditions.  We would like to perform a similar trick in our case, and
regularize the zero-mode infinity by twisting the boundary conditions of the
model to
$\psi^{*\,i}(T) = \exp(i \, \theta_i) \, \psi^{*\,i}(0)$ and
$\chi_i^\mu (T) = \exp(- i \, \theta_i) \, \chi_i^\mu (0)$.  These boundary
conditions respect all the symmetries of the action,
and for non-zero $\theta_i$'s one no longer has zero
modes of the supersymmetries.

For the moment, let us consider a particle moving on a $D$
complex-dimensional target-space torus.  This case is anyway interesting,
and is the only example in which the path-integral can be carried out
explicitly, the gauge-fixed action being quadratic.  Bearing in mind the
transformations of eq\eq{G}, the path integral over a gravitino modulo the
local supersymmetry gives the superJacobian
\beq
\sdet{}_{\theta} \, i \partial_t = \bigl( \det_{\theta} \, i \partial_t
        \bigr)^{-1} \comma \label{Jac}
\eeq
where the index on the ``det'' is to remind us of the shifted boundary
conditions of the fermions.  Thus, integrating over the \einbein{}, the
gravitini and the matter fermions, the partition function on the $D$--torus
reduces to\footnote{We set $\hbar \to 1$ from now on.}
\beq
\calf_{\theta_i} = \frac12 \, \int_0^\infty  \dTTs \,
        \bigl( \det_{\theta_1} \, i \partial_t \bigr) ^{D-1}
        \, \bigl( \det_{\theta_2} \, i \partial_t \bigr) ^{D-1}
        \int \cald X \, \cald \Xbar \, \e^{- \int_0^T d t \;
        G_{\mu\mubar} \, \dot \Xbar{}^\mubar  \dot X ^\mu} \stop
        \label{torus}
\eeq
To evaluate the determinants, one must first square them, to obtain the
positive-definite operator $-\partial_t{}^2$, which has eigenfunctions
$\Psi_{n}^{(\theta)} \sim \e^{i(2\pi n +\theta) \, t/T}$ and eigenvalues
$(2\pi n +\theta)^2/T^2$.  The determinant is proportional to the product
formula for sine's, and one can
fix the
proportionality constant using the zeta-function regularized
result for the periodic case: $\det' \, \left( -\partial_t{}^2 \right) = T^2$
\cite{nicepolch}.
The result is\footnote{Particle theories can have global anomalies.
For example, the
``$N=1/2$'' particle is anomalous unless the target space is a spin manifold
\cite{witanom,AG}.
The fact that $\det_{\theta} \, i \partial_t$ is periodic only
up to a sign is an indication of such an anomaly in \uone--invariant quantum
mechanics theories \cite{israelis}.  In our case there is never any anomaly:
the initial sign of the path integral is ambiguous at the starting point
in field space, but the sign can then be uniquely fixed over the entire space
by considering the case when the two  $\theta_i$'s
are equal---there then being an even number of determinants.}:
\beq
\det_{\theta}^2 \, i \partial_t = - 4 \, \sin^2 \bigl( \theta/2 \bigr)
\onArrow{\theta \to 0} -\, \theta^2 \stop \label{det}
\eeq
If $D > 1$ the path integral
\noeq{torus} vanishes as  we return to periodic boundary conditions.
For the one-torus, it is very reasonable to argue that the regularized
partition function can be defined as the periodic limit of \noeq{torus},
it being completely independent of the
boundary conditions.

At this point we could return to the general problem of the
particle on an arbitrary manifold.  However, the full
evaluation of the partition function on the torus is illuminating in its own
right, so we shall first finish this calculation.  The path integral over
$X$ and $\Xbar$ is standard \cite{nicepolch}.  The usual
constant and non-zero modes give a factor of
$V T/\pi \, \det' \bigl( -\partial_t{}^2 \bigr) = V/T$.  All the
interesting physics comes from the zero modes of $X$.  If one considers a
target-space
torus with complex structure $\sigma$, these are given by all possible windings
of the world-line of the particle around the torus:
\beq
X_{n,m} = ( n + m \sigma ) \, \frac t T \stop
\eeq
(This can be compared to the case of the A model on the torus, for which the
interesting modes are the instantons that exist when the world-sheet and
target-space tori have equivalent complex structures \cite{F1}.)
One thus has
\beqal
\calf &= \frac12 \, \int_0^\infty  \frac \dT {T^{2-s}} \; \frac V \pi \;
         \sumnm \e^{- V \, \frac{|n+ m \sigma|^2}{\sigma_2 T} } \\
      &= \frac12 \, \left( \frac V \pi \right)^s \, \int_0^\infty
         \dTtilde \, \Ttilde^{-s} \; \sumnm
         \e^{- \pi \, \Ttilde \, \frac{|n+ m \sigma|^2}{\sigma_2} }
         \stop \label{int}
\eeqal
To evaluate  the integral, one first splits it into two
regions: $0 \le \Ttilde \le 1$, and $\Ttilde \ge 1$.
Because of the analytic continuation, the large $\Ttilde$ integral is
finite as $s \to 0$.  To examine the behaviour of the integral for small
$\Ttilde$, note that the sum has the form of a heat kernel.
(The reason for this will be clear from the hamiltonian quantization in section
\ref{Hamsec}).  At small times the heat kernel behaves as $1/\Ttilde$ (see
eq\eq{pois}), resulting in a $-1/s$ pole from the integral.  Aside from
this pole, the integral is well behaved as $s \to 0$.  Now,
one can formally interchange the order of the summation
and integration to get
\beq
\calf = \frac12 \, \left( \frac V \pi \right)^s \,  \Gamma(1-s) \;
        \sump \, \left(\frac{\sigma_2}{\pi |n+ m \sigma|^2} \right)^{1-s}
        \stop
\eeq
Then
\beq
\partial_\sigma \calf =  \frac1{4\pi i} \, \left( \frac V \pi \right)^s \,
        \Gamma(2-s) \; \sump \,
        \left(\frac{\sigma_2}{\pi |n+ m \sigma|^2} \right)^{-s} \,
        \frac1{(n+ m \sigma )^2}
        \stop \label{G2}
\eeq

These expressions have several interesting features, which can be
understood from basic principles, and which in fact allow one to evaluate
$\partial_\sigma \calf$ without having to do any calculation.
First, while $\calf$ has a residual dependence on $V$, because of the
pole as $s\to 0$,
$\partial_\sigma \calf$ depends only on the complex structure.
$\calf$ is explicitly modular invariant, as it should be.
This means that $\partial_\sigma \calf$ transforms under modular
transformations with weight 2 (in our normalization). $\calf$ diverges as
$\sigma_2 \to \infty$---this is the mirror of the large volume
behaviour \cite{F1} in the A model---but the divergence is soft enough to
make $\partial_\sigma \calf$ finite in this limit.  Finally, taking the
limit $s \to 0$ naively in \noeq{G2}, one sees that $\partial_\sigma
\calf$ appears to be holomorphic.
In fact these features
are mutually incompatible, since they would mean that
$\partial_\sigma \calf$ would be a holomorphic
modular form of weight 2, and no such
object exists \cite{somebook}!  Clearly, because of the BRST anomaly in the
theory \cite{F1}, what must give way is the
holomorphicity of $\partial_\sigma \calf$.  In fact, the regularized
summation in \noeq{G2} gives the nonholomorphic quantity \cite{somebook}
\beq
\hat G_2(\sigma) = - 4 \pi i \, \partial_\sigma \log  \eta(\sigma)
                   - \frac\pi{\sigma_2} \stop
\eeq
Integrating \noeq{G2}, recalling the pole term in \noeq{int}, and using the
symmetry between $\sigma$ and $\sigmabar$ in \noeq{int}, we
find\footnote{In their evaluation of threshold corrections in string
theories, Dixon, Kaplunovsky and Louis essentially calculated the one-loop
partition function $\calf_1$ of the $N=2$ string on a one-torus
\cite{kapl}.  Their $\calf_1$ is symmetric between the \Kahler{} and
complex structure---a reflection of mirror symmetry---and can be regarded as
the full partition function of both the A and the B sigma
models. Eq\eq{int} corresponds to the case of their ``degenerate maps''.}
\beq
\calf = - \,\frac12 \, \log \left( \, V \, \sigma_2 \,
        \eta^2(\sigma) \, \eta^2(\bar\sigma) \, \right) \comma \label{fans}
\eeq
up to an infinite additive constant.  The complex
structure dependence of \noeq{fans} agrees with  BCOV
\cite{F1}.

We have seen that on the one-torus the anomalous dependence of
$\partial_\sigma \calf$ on $\bar\sigma$  can be traced to a conflict between
modular covariance and holomorphicity, and that $\partial_\sigma
\calf$ can be determined uniquely in this case using only general properties.
It would be interesting if this
interpretation of the holomorphic anomaly could be generalized to other
target spaces, but the theory of modular forms on \CY{} spaces is apparently
undeveloped.

\section{Hamiltonian quantization and the \RS{} torsion}
\label{Hamsec}

After this long digression, we can return to the problem of finding $\calf$ on
a general target space.  We have argued that one can regulate the infinity
coming from the volume of the space of constant supersymmetries by examining
the theory with twisted boundary conditions.  The \einbein{} and the gravitini
can then be completely gauged away, reducing the problem to that of calculating
the partition function of the ungauged particle \noeq{nograv} on a circle
with period $T$; the only remnant of the supergravity fields being that one
should integrate over $T$ with the measure \noeq{einb}, and that one should
insert the regulated Jacobian coming from \noeq{Jac} and \noeq{det}.  The
partition function on the circle can then be calculated in the hamiltonian
formalism, giving
\beq
\calf = - \frac12 \; \lim_{\theta_i \to 0} \, \int_0^\infty  \dTTs \;
        \frac1{\theta_1\theta_2} \, \Tr \, (-1)^F \e^{i \theta_1 F_1}   \,
        \e^{i \theta_2 F_2}\, \e^{-H \, T} \stop
\eeq
Here $H$ is the hamiltonian corresponding
to eq\eq{nograv}, and
the twisted boundary conditions on the spinors $\chi_i^\mu$ are
implemented by the insertion of $\e^{i \theta_i F_i}$, $F_i$ being the
appropriate fermion number operator.  Expanding in the $\theta$'s, in
order to carry out the limit, the leading term proportional to the
Euler number is regulated to zero.  The $1/\theta_i$ terms vanish by
CPT, so one is left with our final expression:
\beq
\calf = \frac12 \, \int_0^\infty  \dTTs \;\,
        \Tr \, (-1)^F \, F_1  \, F_2 \; \e^{-H \, T} \stop \label{done}
\eeq

This expression is clearly very similar to the index $\calf_1$ of \noeq{F1},
which was postulated by BCOV to be the one-loop
partition function of the B sigma model coupled to
topological gravity in \cite{F1}.  However, it should be borne in mind
that  $\calf_1$ is defined over the Hilbert space of the string, whereas
$\calf$ is defined over the much smaller Hilbert space of the particle.
In \cite{KS} $\calf_1$ was evaluated in the B model by
considering the large volume limit, and was seen to be related to the \RS{}
torsion of the manifold.  In our case, we can evaluate
$\calf$ directly.

There are several issues that need to be addressed in the hamiltonian
quantization of the theory.  The first is the choice of canonical variables:
following Witten \cite{B}, we would like $\Q 1$,  ``the BRST operator'' of the
theory, to act as the Dolbeault operator.  In order to do this,
it turns out to be necessary to drop the manifest \utwo{} symmetry of the
action, and to take as the canonically conjugate variables:
$$
\begin{array}{ccc}
X^\mu & \leftrightarrow & P_\mu \\
\Xbar^\mubar & \leftrightarrow & \Pbar_\mubar \\
\rho_\mubar \equiv \chi_{1\mubar}& \leftrightarrow &
        \eta^\mubar \equiv \chi^{*\,1\mubar} \\
\rhotilde^\mu \equiv \chi_{2}^\mu & \leftrightarrow &
         \theta_\mu \equiv \chi_\mu^{*\,2}  \stop
\end{array}
$$
Next one has to fix some operator orderings:
The supersymmetry charges $\Q i$ and $\Qb i$ and the hamiltonian
$H$ are determined classically by varying the gauged action
\noeq{particle} with respect to the supergravity fields.  In order for the
$(2,2)$ supersymmetry algebra to close, these operators must be ordered as
\beqaltwo
&\Q1 = \eta^\mubar \Pbar_\mubar
&&\Qb1 = G^{\mu\mubar} \,  \rho_\mubar
        \left( P_\mu + i \, \Gamma_{\mu\rho}^\sigma \, \rhotilde^\rho
        \theta_\sigma \right) \nonumber\\
&\Q2 = G^{\mu\mubar} \,  \theta_\mu
        \left( \Pbar_\mubar + i \, \Gamma_{\mubar\rhobar}^\sigmabar
        \eta^\rhobar \rho_\sigmabar \right)
&&\Qb2 = \rhotilde^\mu P_\mu   \comma \label{ops}
\eeqaltwo
with
\beq
H = \left \{ \Q1 \, , \, \Qb1 \right \}
   = \left \{ \Q2 \, , \, \Qb2 \right \} \stop \label{ham}
\eeq
This ordering leads us to a different interpretation for the
Hilbert space of the particle than that of Witten's for the B model
\cite{B}.  He took
$\eta^\mubar$ and $\theta_\mu$ to be creation operators, so that in the
large-volume limit, the states of the theory were equivalent to the space of
$(0,q)$ forms in $\wedge^p \,T\,(M)$, or simply the antisymmetric tensors
$A_{\mubar_1 \cdots \mubar_q}^{\mu_1 \cdots \mu_p}$'.
Here we see that for $\Q1$ to
represent $-i \pb$, $\eta^\mubar$ should indeed be a creation
operator.  However, for $\Qb1$ to then be the geometrical operator
$-i \pb^\dagger$, $\rhotilde^\mu$ and not $\theta_\mu$ must be the other
creation operator!  (The theory is then symmetrical under combined complex
conjugation and multiplication by $\sigma_1$.)
This means that {\em the B model is a theory with
operators acting on the $(p,q)$ forms of a \Kahler{} manifold}.  Of course,
on a \CY{} manifold one can always use the holomorphic tensor
$\Omega_{\mu_1 \cdots \mu_n}$ to convert between holomorphic vectors and
forms.

With this interpretation of the fermionic operators, $F_1$ and $F_2$ simply
give $q$ and $p$, respectively\footnote{The $F$'s are determined only
up to signs and additive normal ordering constants.  Since the
analogue of $\calf$ without both fermion number insertions vanishes, these
ambiguities can change  $\calf$ only by an sign.  As we
noted previously, the signs of the the fermionic determinants used in
the derivation of $\calf$ are also ambiguous, and we determine the
overall sign of $\calf$ by comparison to the lagrangian result \noeq{int}.},
and the hamiltonian of the theory becomes
$\{ -i \pb \,,\, -i \pb^\dagger \} = - \nabla^2_{p,q}$ ---the
(negative of the) laplacian acting on the $(p,q)$ forms of the
target space.  Thus
\beqal
\calf &= \frac12 \, \sum_{p,q} \; (-1)^{p+q} \, p \, q \;
        \int_0^\infty  \dTTs \; \Tr \; \e^{\nabla^2_{p,q} \, T} \\
      &\equiv \, \frac12 \, \sum_{p,q} \; (-1)^{p+q} \, p \, q \;
        \Gamma(s) \; \zeta (s,  \nabla^2_{p,q}) \comma \label{\RS}
\eeqal
where we have introduced the zeta-function of the laplacian $\zeta (s,
\nabla^2_{p,q})$.
If $\nabla^2_{p,q}$ had no zero modes, one would have
$\zeta(0,\nabla_{p,q}^2)=0$, so that
$\Gamma(s) \, \zeta(0,\nabla_{p,q}^2) \to \zeta'(0,\nabla_{p,q}^2)$.
$\calf$ would then be precisely the sum, weighted by $(-1)^p \, p$, of the
logarithms of the \RS{} torsions $\log T_p$ \cite{\RS}.
With zero modes one has extra infinite and anomalous
pieces, as we saw in the torus case in \noeq{fans}, but these drop
from $\pa_\sigma \calf$.  (Note that
the zeta-function regularization is crucial in the formula \noeq{\RS}
for the \RS{} torsion.  $\calf$ is formally the sum of the
logarithms of the eigenvalues of $H$, but this sum is highly divergent.
This will also be true for the string.  In a field theory approach,
dimensional regularization will play the role of the
zeta-function regularization.)
When the target space is a \CY{} manifold, BCOV argued that their index
$\calf_1$ is also given by \noeq{\RS} (up
to regularization issues) \cite{F1}.
Having obtained the same answer, we have an
\apost{} justification of the formal arguments we used in  deriving the
particle action \noeq{particle}.

Returning again to the one-torus, we can now
explicitly evaluate $\calf$ in the hamiltonian formalism.  The
eigenvectors $H$ are
$$
\Psi_{n,m} = \frac1{\sqrt{\sigma_2}}
  \; \e^{\pi \, \frac{n+m\sigmabar}{\sigma_2} \, z} \;
     \e^{- \pi \, \frac{n+m\sigma}{\sigma_2} \, \zbar} \comma
$$
with eigenvalues
$$
\frac{\pi^2}V \, \frac{|n+ m \sigma|^2}{\sigma_2} \stop
$$
Substituting these into  \noeq{\RS}, and noting that $p$ and $q$ range from 0
to 1, one gets
\beqal
\calf &= \frac12 \, \int_0^\infty  \dTTs \;
         \sumnm \e^{- \frac{\pi^2 T}V \, \frac{|n+ m \sigma|^2}{\sigma_2} } \\
      &= \frac12 \, \left( \frac V \pi \right)^s \, \int_0^\infty
         \dTT\, T^s \;
         \sumnm \e^{- \pi \, T \, \frac{|n+ m \sigma|^2}{\sigma_2} }
         \comma \label{hamint}
\eeqal
to be compared to our previous expressions in \noeq{int}.
To evaluate \noeq{hamint} one
needs the Poisson resummation formula for the heat kernel \cite{\RS}:
\beq
\sumnm \e^{- \pi \, T \, \frac{|n+ m \sigma|^2}{\sigma_2} } = \frac1T
\; \sumnm \e^{- \frac\pi T \; \frac{|n+ m \sigma|^2}{\sigma_2} }
         \excl \label{pois}
\eeq
(This somewhat surprising formula might have been expected
from target-space modular invariance.)  As a
result, one finds that the integral in
\noeq{hamint} is invariant under the interchange $s \leftrightarrow (1-s)$,
giving a typical zeta-function identity
$\Gamma(s) \, \zeta (s,\nabla_{p,q}^2)  =\Gamma(1-s) \,
\zeta(1-s,\nabla_{p,q}^2)$.  Using this identity,
\noeq{hamint} agrees perfectly with the Lagrangian calculation of \noeq{int}.

Finally, hamiltonian quantization in the open case leads to the immediate
analogue of \noeq{\RS}
\beq
\calf = \frac12 \, \sum_{q} \; (-1)^{q} \, q \;  \Gamma(s) \;
        \zeta (s,\nabla_{0,q}^2) \stop \label{\RS o}
\eeq
This theory becomes much more interesting if one introduces background gauge
fields, by inserting Wilson lines in some representation $R$ of the
Chan--Paton group into the path integral.  The string carries group indices
on both of its ends, so string states are in (some subset of) the $R \otimes
R$ representation.  Particle states are simply in $R$.  In general, the path
integral with Wilson lines cannot be easily evaluated.  However, when one
implements the Wilson loop insertions locally in the particle using boundary
fermions \cite{boundfer}, a hamiltonian quantization of the theory shows that
one simply replaces the laplacian operator by the appropriate covariantized
laplacian.  It is important that, as was pointed out by Witten for the B
string, one can introduce only gauge fields for which the associated field
strength has a vanishing $(0,2)$ component \cite{CS}.  This means that the
states of the theory are in a {\em holomorphic\/} vector bundle $E$.  The
\RS{} torsion of the bundle can then be defined, and the partition function
is given by $\calf = \log T_0 \, (E)$.

\section{Deforming the \Kahler{} and complex structure of the manifold}

So far, we have considered the particle defined on a manifold with a fixed
\Kahler{} and complex structure.  It is interesting to also consider how the
theory can be deformed.  In \cite{B} the topological B model was varied using
the two-form operators in the BRST cohomology of the model.  Later, this was
generalized to include ``anti-topological'' and ``mixed'' deformations (see
\cite{KS}).  We do not wish to use these arguments, which are based upon how
one varies topological or twisted $N=2$ superconformal theories.  Instead,
we simply
look for all transformations of the theory that preserve the supersymmetry
algebra.  Since $(2,2)$ sigma models can only be defined on \Kahler{} spaces,
this reduces to the question of how to deform \Kahler{} spaces.

The simplest  deformation is to keep the complex structure
fixed, and to change the \Kahler{} metric.  Such a deformation can be carried
out explicitly in the particle.  Under an infinitesimal
change $g_{\mu\mubar} \to g_{\mu\mubar} + h_{\mu\mubar}$, one sees from
\noeq{ops}
that $\Q1$ and $\Qb2$ are clearly invariant, whereas $\Qb1$ and $\Q2$
change by
\beqal
\delta \Qb1 &= \left [  \Qb2 \, , \, X \right ] \nonumber \\
\delta \Q2 &= - \left [  \Q1 \, , \, X \right ] \comma \label{delk}
\eeqal
with
\beq
X = h^{\mu\mubar} \rho_\mubar \, \theta_\mu \stop \label{h}
\eeq
In order for the $N=2$ supersymmetry algebra to be preserved, X must
commute with $\Qb1$ and $\Q2$.  This happens {\em iff\/} the 2--form
$h_{\mu\mubar}$ is closed which, of course, is the case for a deformation
of the \Kahler{} form.  Using \noeq{ham} and \noeq{delk}, one sees
that the hamiltonian is changed by
\beq
\delta H = \left \{ \Q1 \, , \,
                \left [  \Qb2 \, , \, X \right ] \right \}
                        \stop \label{varh}
\eeq
Note that whereas in the sigma model one has a {\em
complexified\/} \Kahler{} structure, so that there are two independent
variations like \noeq{delk} and
\noeq{varh}, here there is only one such variation.

The only other variation one can make on a \Kahler{} space is to deform
its complex structure.  This gives a finite-dimensional space of
transformations.  They are harder to carry out explicitly in the particle,
since the complex structure
appears only implicitly in the Lagrangian \noeq{particle} and the
supersymmetry operators \noeq{ops}.  However, knowing that under changes of
complex structure $\pa$ and $\pb$ mix, one is led, in analogy to
\noeq{delk}, to try the transformations:
\beq
\Q1 \to \Q1 - \left [  \Qb2 \, , \, Y \right ] \qquad
\Q2 \to \Q2 + \left [  \Qb1 \, , \, Y \right ]   \label{delc} \comma
\eeq
with the $\Qbar$'s unchanged.  Because of the equality of the two expressions
for $H$ in \noeq{ham}, the same $Y$ must appear in both transformations.
One also has the complex conjugate
transformations:
\beq
\Qb2 \to \Qb2 + \left [  \Q1 \, , \, \Ybar \right ] \qquad
\Qb1 \to \Qb1 - \left [  \Q2 \, , \, \Ybar \right ]
        \comma \label{delcb}
\eeq
with the $Q$'s unchanged.  By counting dimensions and the two fermion
numbers, one sees that $Y$ and $\Ybar$ take the form
\beqal
Y &= A_\mubar^\mu \, \eta^\mubar \theta_\mu \nonumber \\
\Ybar &= \Abar^\mubar_\mu \,  \rho_\mubar \rhotilde^\mu \comma \label{Y}
\eeqal
with $A_\mubar^\mu$ and $\Abar^\mubar_\mu$ some tensors on the target space.

We shall now concentrate on the $A_\mubar^\mu$ transformations; the case of the
$\Abar^\mubar_\mu$ transformations follows by complex conjugation.
In order for the modified $\Q1$ to be nilpotent, one sees that,
infinitesimally, $\pb A = 0$.  This condition is Witten's statement that
the $Y$ in \noeq{Y}
should be a zero-form in the BRST cohomology of the theory, which he
identified as generating (holomorphic) changes of the complex structure.
Under the transformation  \noeq{delc}, the hamiltonian changes by
\beq
H \to H + \left \{ \Qb2 \, , \,
        \left [  \Qb1 \, , \, Y \right ] \right \} \comma \label{delh}
\eeq
which is the two-form operator corresponding to $Y$.
In fact, it was noticed in
\cite{lamba2} that $\Q1$ remains nilpotent under {\em finite\footnote{Note,
however, that the transformations
\noeq{delc} and \noeq{delcb} are incompatible.  Thus, one can only have a
finite holomorphic deformation if one keeps the antiholomorphic part of the
complex structure fixed.  This agrees with the well-known result that
the space
of complex-structure deformations is not affine.  Using \noeq{delc} for
finite transformations means that we are using ``canonical
coordinates'' \cite{KS} on the moduli space of complex structures.}\/}
transformations \noeq{delc}, as long as $A_\mubar^\mu$
satisfies the full \KS{} equation for the variation of complex structures:
\cite{\KS}
\beq
\pb A^\mu + A^\nu \pa_\nu A^\mu = 0 \stop \label{\KS}
\eeq
(This equation has been written thinking of $A^\mu$ as a one form.  If
one writes $A$ as well as a vector
field, \noeq{\KS} can be written more geometrically as
$\pb A + 1/2 \; [  A \, , \, A ]  = 0 $.)  In view of this, we see that
eqs\eq{delc} with \noeq{Y} do indeed
represent a (finite) holomorphic change of the complex structure of the
manifold.  We have therefore found all the possible deformations of the theory.

It is still necessary to check that the modified theory satisfies the
full $(2,2)$
supersymmetry algebra.  It is easy to see that all the anticommutators
involving  $\Qb i$'s close, taking into account the fact that the hamiltonian
has been  modified \noeq{delh}.  The remaining conditions are that $\Q2$ be
nilpotent, and that it anticommute
with $\Q1$.  Repeatedly using the Jacobi identity, one finds that
this occurs if $A$ satisfies the auxiliary condition
\beq
\cald^{[\,\nu} A_\mubar^{\mu \,]} +
        A_\rhobar^{[\,\nu} \cald^\rhobar A_\mubar^{\mu \, ]} = 0
        \label{extra} \semi
\eeq
here the brackets indicate antisymmetrization, and $\cald$ is the (raised)
covariant derivative on the manifold.  The geometrical reason for this
condition is that in a \Kahler{} space $\{\pb,\pb^\dagger\} =
\{\pa,\pa^\dagger\} = \nabla^2$,
so complex-structure deformations of $\pb$ and
$\pa^\dagger$ must be related.  Eq\eq{extra}  states that the
deformed $\pa^\dagger$ is nilpotent and anticommutes with $\pb$.  There is
also clearly a nice symmetry between the \KS{} equation and \noeq{extra}.
However, we have not encountered this equation in the literature.  (Of course
it would  have arisen in the discussion on the B sigma model in
\cite{B,lamba2} if the extra BRST operator of the theory would have been
considered.)

In general, equations like the \KS{} equation and
\noeq{extra} are difficult to solve, and it is even difficult to know when
they have solutions.  We have  been able to show that {\em in the
infinitesimal case\/}, one can use the diffeomorphism invariance of the
theory $A \to A + \pb \, \xi$ to solve \noeq{extra} if the manifold is \CY{}
or if $H^{(0,2)}$ of the manifold vanishes\footnote{We would like to thank Ori
Ganor for  helpful discussions on this point.}.  However, from our
derivation, whenever there is a complex-structure deformation on a \Kahler{}
manifold, one should {\em always\/} be able to represent it by a solution
$A$ of the \KS{} equation that also satisfies \noeq{extra}.

\section{Independence of the particle on the \Kahler{} structure}

Using BRST invariance, Witten argued that the B model should not depend on
the \Kahler{} structure of the target space.  (This is also essentially true
of the \RS{} torsion \cite{\RS}.)
In \cite{KS}, BCOV showed (for
genus $g > 1$) that this result remains true
despite the presence of BRST anomalies\footnote{At genus one, the calculation
  is complicated by the fact that one has to work with derivatives of
  $\calf_1$, rather than with $\calf_1$ itself.}.
Using the results of the previous section, we can reproduce the argument of
BCOV for the simpler case of the particle.
First, consider the variation of $\calf$ under an
infinitesimal change of complex structure \noeq{delc}.  Expanding the
``0--form'' generator  $Y$ in \noeq{Y} as $Y = t^i \, \Yi$, and
substituting the variation of the
hamiltonian \noeq{delh} into our expression for $\calf$ \noeq{done}, one gets
\beq
\pa_{t^i} \calf = - \, \frac12 \, \int_0^\infty  \dT \, T^s \;
        \Tr \; (-1)^F \, \Qb1 \, \Qb2 \, \Yi  \; \e^{-H \, T} \stop
        \label{dF}
\eeq
Unlike $\calf$ itself, $\pa_{t^i} \calf$ is finite and well-defined,
and since the $\Qbar$'s come from the $G$'s of
the sigma model, it has the traditional form of a
topological one-point function on a torus \cite{DVV}.

We would now like to deform the \Kahler{} metric in \noeq{dF}.  Recalling
from \noeq{delk} that this is done by changing $\Qb1$ and $H$, with
$\delta \Qb1 = \left [  \Qb2 \, , X \right ] $,
one obtains
\beqal
\pa_{t^i} \, \delta \calf = - \, \frac12 \, \int_0^\infty  & \dT \, T^s \;\;
        \Tr \; (-1)^F \, \cdot \;
        \Bigl( \, \Qb2 \, X \, \Qb2 \, \Yi  \; \e^{-H \, T} \\
        & - \int_0^T  \dt \; \Qb1 \, \Qb2 \, \Yi \; \e^{-H \, t}
        \left \{ \Q1  , \left [  \Qb2 \, , X \right ] \right \}
        \; \e^{-H \, (T -t) } \, \Bigr) \stop \label{mess}
\eeqal
Our calculation is now  similar to that of the holomorphic anomaly in
\cite{ising},  except that our quantities are nicely regularized.  First, we
move $\Q 1$ in the second term around the trace.  Since $\Yi$
is in the ``BRST cohomology'', $\Q1$ moves through everything except
for $\Qb1$, with which it anticommutes to give a factor of $H$.
This gives us
\beqal
\pa_{t^i} \, \delta \calf &=  \frac12 \, \int_0^\infty  \dT \, T^s \;
        \frac{{\rm d}}{\dT} \; \Tr
        \int_0^T  \dt \; (-1)^F \Qb2 \, \Yi \; \e^{-H \, t} \;
        \Qb2 \, X \; \e^{-H \, (T -t) } \\
&= - \, \frac s2 \, \int_0^\infty  \dTTs
        \int_0^T  \dt \; \Tr \; (-1)^F \Qb2 \, \Yi \; \e^{-H \, t} \;
        \Qb2 \, X \; \e^{-H \, (T -t) } \label{mess2} \stop
\eeqal
Because of the explicit factor of $s$, $\pa_{t^i} \, \delta \calf$ vanishes
unless the $T$
integration gives a pole as $s \to 0$.  This will occur {\em iff\/}
the $t$ integral is finite and nonzero as $T \to 0$ or $T \to \infty$.

As $T \to \infty$, being careful to keep all contributions, the
integral tends to
\beq
\int_0^{T/2}  \dt \; \Tr \; (-1)^F \,
        \left( \,  \Qb2 \, \Yi  \; \e^{-H \, t} \, \Qb2 \, X P
        + \Qb2 \, \Yi  \, P \, \Qb2 \, X \; \e^{-H \, t} \, \right)
        \label{mess3} \comma
\eeq
$P$ being the projection onto the
$H=0$ sector of the theory.
This sector is supersymmetric, so it is annihilated by $\Qb2$.  Cycling
$P$ to be next to $\Qb2$, one sees that \noeq{mess3} vanishes,  so there is
no contribution to
$\pa_{t^i} \, \delta \calf$ from the $t$ integral at large $T$.
For small $T$, it appears to be clear that the $t$ integral vanishes.  The
only subtlety is that the heat kernel diverges as $1/T^d$ for small
times---$d$ being the {\em complex} dimension of spacetime.  (Such
divergences gives rise to the contact terms that appear in the string
derivation of the anomaly \cite{F1,KS}.)  Using the fact that the product of
a local operator times the heat kernel can be written as a Laurent series in
$T$ \cite{Gilkey}, and knowing that $\pa_{t^i} \, \delta \calf$ is
finite, one can see that the integral indeed vanishes in this limit.  (This
can also be checked explicitly in the torus case.)  Thus $\pa_{t^i} \, \delta
\calf=0$.

A similar argument shows that $\pa_{\tbar_i} \, \delta \calf$ also vanishes.
Thus, as we saw in the torus case \noeq{fans}, the partition function of the
particle depends on the \Kahler{} structure of the metric only by a trivial
additive factor, independent of the complex structure.

\section{The holomorphic anomaly}

The holomorphic anomaly is derived in a very similar manner.  One now wants
to substitute the antiholomorphic variation of the complex structure of the
$\Qbar$'s and of $H$ coming from eq\eq{delcb} into
$\pa_{t^i} \, \calf$\footnote{In \cite{ising} and \cite{F1} the terms coming
from varying the $\Qbar$'s were not considered.  As we have argued, they
should be important in shifting the ``basepoint'' of the theory.}.
This gives
\beqal
 \pb_{\tbar_\ibar} \pa_{t^i} \, \calf = - \, \frac12 \,
        \int_0^\infty  & \dT \, T^s \;\; \Tr \; (-1)^F \, \cdot \; \Bigl( \, \\
        & \left( \Qb1 \, [ \, \Q1 \, , \, \Ybi \, ] \, \Yi
        +  [ \, \Ybi \, , \, \Q2 \, ] \, \Qb2 \, \Yi \, \right)
        \; \e^{-H \, T} \\
        & - \int_0^T  \dt \; \Qb1 \, \Qb2 \, \Yi \; \e^{-H \, t}
        \left \{ \Q2  , \left [  \Q1 \, , \Ybi \, \right ] \right \}
        \; \e^{-H \, (T -t) } \, \Bigr) \stop \label{newmess}
\eeqal
Again cycling the $Q$'s around the trace and integrating by parts,
this simplifies to
\beqal
 \pb_{\tbar_\ibar} \pa_{t^i} \, \calf &=
        - \, \frac s2 \, \int_0^\infty  \dTTs \, \cdot \; \\
    & \qquad  \qquad
       \Tr \; (-1)^F  \,  \Bigl( \, \Ybi \; \Yi \; \e^{-H \, t} \;
       - \, \int_0^T  \dt \; H \; \Yi \; \e^{-H \, t} \;
        \Ybi \; \e^{-H \, (T -t) } \Bigr) \\
    &= - \, \frac{s \, (1-s)}2 \, \int_0^\infty  \dTTTs \,
       \int_0^T  \dt \; \Tr \; (-1)^F  \, \Yi \; \e^{-H \, t} \;
        \Ybi \; \e^{-H \, (T -t) } \label{newmess2} \stop
\eeqal
Again the explicit $s$ factor can be canceled only from a logarithmic
divergence in the $T$ integral.  This means that the anomaly is given by
one-half of the coefficient of the term in the $t$ integral linear in
$T$, evaluated between infinity and zero.  (Equivalently, one-half of the
constant in $T$ piece of the second line of \noeq{newmess2}.)
As $T \to \infty$, one gets
\beq
 \pb_{\tbar_\ibar} \pa_{t^i} \, \calf = \frac 12 \,
    \Tr \; (-1)^F  \, \Yi \; P \; \Ybi \; P \comma
\eeq
which gives us the easier part of the anomaly.

The contribution to the anomaly from $T \to 0$ is again subtle.  In the
closed string case, it comes from contact interactions between the analogues
of $\Yi$ and $\Ybi$.  Here, it should come from the small-time expansion of
the heat kernel.  Using the same sort of argument we had previously,
one sees that one can get a finite
contribution to the anomaly from the constant terms in the small-time
expansions of
$\Yi \; \e^{-H \, T}$ and $\Ybi \; \e^{-H \, T}$ in \noeq{newmess2}.
Unfortunately, we have not
not yet been able to  calculate these terms.  We do, however, know the
answer!  As we have noted before, the partition function is the sum,
weighted by $(-1)^p \, p$, of the
logarithms of the \RS{} torsion $\wedge^p T^*$.
The holomorphic (Quillen)
anomaly of the \RS{} torsion of any holomorphic vector bundle $V$ has been
calculated \cite{BF}, giving
\beq
\pa \, \pb \; {\rm log} \, T(V) =
 \frac12 \, \pa \, \pb \sum_q (-1)^q \ d_q
 +\pi i\int_M {\rm Td}(T)\,  {\rm Ch}(V) \, \Big |_{\, (1,1)} \stop
\eeq
(See \cite{KS} for an explanation of the symbols.)
Therefore the small $T$ behaviour of \noeq{newmess2} must give
\beq
- \pi i \int_M  {\rm Td}(T) \, \sum_p \, (-1)^p \, p \;  {\rm Ch} \,
          ( \wedge^p \, T^*) \stop \label{math}
\eeq

\section{Conclusions}

Witten showed that the partition function of the B model coupled to
topological gravity is dominated by world-sheets that collapse to
one-dimensional ``nets'' \cite{CS}.  Since these nets can be produced by
Feynman diagrams, he argued that one should be able to describe the B models
by ordinary field theories.  It is obvious that on such world sheets the
sigma model dimensionally reduces to a one-dimensional sigma model.  The
difficulty is how to represent the coupling to the topological gravity.  Here
we have suggested that this coupling should be replaced by a twisted coupling
of the sigma model to a one-dimensional supergravity.  Thus the topological
sigma model is reduced to the supersymmetric
twisted spinning particle of eq\eq{particle}.  By performing a hamiltonian
quantization of the theory one can see explicitly---after clearing up a few
subtleties---that the one-loop partition function of the spinning particle
\noeq{done}
agrees with that of the B string.  Note that because the particle can
be defined for any complex structure, the partition function is more
like a generating function, since it effectively describes all the
amplitudes of the theory at one loop.
The particle has the advantage that, at
least at one loop, it gives a more general description of the B model than
the two-dimensional sigma model approach.  This is because there are no local
anomalies in particle theories, so the particle can be defined on any
\Kahler{} manifold, and not only when the \CY{} condition is satisfied.  It is
natural to speculate that the particle on a non-\CY{} manifold is the mirror
of a nonconformal topological A model.

A somewhat surprising feature that arises from the hamiltonian quantization
is that the Hilbert space of the particle is naturally described by $(p,q)$
forms.  This is despite the fact that the BRST cohomology of the B string
corresponds to $(0,q)$ forms in $\wedge^p \,T\,(M)$ \cite{B}.  We may note
that in particle theories one does not have a one-to-one correspondence
between allowed deformations and states in the theory.  Thus the deformations
of the theory correspond to changes of the complex structure of the \Kahler{}
manifold, and are indeed described by
the tensors $A_\mubar^\mu$'s, and their complex conjugates.

While from a physics viewpoint we would be disappointed if the particle could
not be generalized to give other amplitudes, from the mathematical point of
view it is already interesting that we can write a particle model that can
describe \RS{} torsion on a manifold.  (One would hope that the anomaly
derivation in section 8 can be completed, so that the particle description
would already show some practical use at this stage.)
This is analogous to the one-dimensional
sigma models written to calculate index theorems \cite{AG,Paul}.
The partition function of the B model is written in terms of a particular sum
of the \RS{} torsion over holomorphic $p$--forms \cite{KS}.  One can find
interesting alternative theories by gauging various subgroups of the global
\utwo{} symmetry of the particle \cite{to appear}.  In particular, if one
gauges the \uone{} generated by $(1-\sigma_3)$, and adds an appropriately
normalized \CS{} term \cite{israelis}, one can get a description of the \RS{}
torsion on $\wedge^p \, T^*$ for any particular $p$.  We have also
seen that in the particle description of the open string---obtained by
truncating \noeq{particle} to a \uone{} action---one can couple
the theory to appropriate background gauge fields \cite{CS} to get the \RS{}
torsion of $E$, a holomorphic vector bundle in some representation of the
gauge group.

We have stressed that the holomorphic anomaly on the one-torus
can be seen as arising from a conflict between holomorphicity and modular
invariance, and have speculated about a possible generalization of this to
other manifolds.  We have also derived an auxiliary condition to the \KS{}
equation, which appears to be necessary for variations of the complex
structure of a \Kahler{} manifold.  It is not clear to us whether or not this
last result is surprising or obvious.

\section{Comments on possible field theories}

So far, all our discussion has been of the particle at one loop.  In that
case the only remnants of the gravitini are the insertions of the fermion
number operators $F_i$ in the partition function \noeq{done}.  Therefore one
might suspect that the success of the particle in reproducing the one-loop
partition function does not necessarily imply that the particle action
\noeq{particle} is correct.  One indication that we have the right coupling
to gravity is that at higher loop amplitudes one would expect
``zero-modes'' of the gravitini to give rise to insertions of the local
supersymmetry currents, as  occurs in two-dimensional theories
coupled to topological gravity \cite{DVV}.   These currents in
the particle are those appropriate to the B model.

If one wants to proceed to arbitrary amplitudes, one will have to do one of
two things: either to calculate the path integral of the particle on nets, or
to write the relevant field theory.  At least in principle it is easy to
calculate string amplitudes on complicated Riemann surfaces.  In particle
theories one has the fundamental problem of having to introduce interactions
at the vertices.  In our case we do have a very natural geometrical candidate
for such an $n$--point vertex: one simply takes the forms corresponding to
the states of the $n$ particles, and integrates their product over the
manifold.  In general, one might expect that only the 3-point vertex will be
needed in the theory, since the theory is at least somewhat topological (and
since the \KS{} equation is quadratic).  However, it is not obvious that this
will be the case, and the only check will be to see whether the interacting
particle can generate the genus $g$ holomorphic anomaly of BCOV \cite{KS}.

In general, it is not easy to find a string field theory from a string.
One first needs to know the space of the string
fields, then the theory's kinetic operator and finally its
interactions.  Normally one knows that the string field should describe the
space of states of the first quantized gauge-fixed string, including its
ghosts.  The linearized equations of motion and gauge invariances of the
field theory then come from the BRST operator of the string.  Although such
an approach was very successful in describing the open string \cite{OS},
there can be complications to this method,
as witnessed in the difficulties in the
construction of the closed-string field theory \cite{Barton}.  An alternative
approach is to note that the beta-functions of the string coupled to
background fields should be the low-energy equations of motion of the field
theory.  In the more topological theories, such as the various $N=2$ strings
\cite{N=2} and the B strings \cite{CS,KS}, one may hope that these equations,
which for some reason are always quadratic,
could be exact.

In the case of the particle, things are even more difficult.  As we have
already stated, the first problem is that the above two approaches do not
match.  The Hilbert space of the particle is the space of forms, and the
constraints  from varying the supergravity fields in \noeq{particle}
are that $\pa$, $\pb^\dagger$ and $\nabla^2$ all vanish on these forms.  This
means that the BRST cohomology in the Hilbert space of the particle should be
equivalent to the de Rham cohomology of the target space, although things
might be more complicated because of the commuting supersymmetry
ghost system.  On the other hand, the deformations of the theory are
described by the fields
$A_\mubar^\mu$ satisfying the \KS{} equation \noeq{\KS} and our auxiliary
equation \noeq{extra}.  These spaces are only
compatible on \CY{} manifolds.  An alternative statement of the problem is
that in string theories the legitimate conformally invariant vertex operators
are (generally) equivalent to the states in the BRST cohomology.  In particle
theories, there is no constraint from conformal invariance, and one has a
free choice of vertex operators \cite{polybook}.

With all these caveats, the most obvious possibility for a field theory of
the B string is still to take the $A_\mubar^\mu$'s as the basic fields of the
theory, and to choose an action whose equation of motion gives the \KS{}
equation.  This gives the ``\KS{} field theory'' of BCOV \cite{KS}.  However,
having the correct classical equations of motion may not be enough to fix the
full field theory, and one must be very careful to know that one is working
on the correct Hilbert space.  As an example of this, the $N=2$ closed string
has an analogous ``Plebanski action'' \cite{N=2}.  However, there appears to
be a discrepancy between the one-loop three-point function calculated from
the string, and that calculated from the field theory \cite{italians}.
The fact that the
\KS{} action in \cite{KS} is nonlocal may be a warning sign that one does not
yet have the correct Hilbert space.  The final check will again be whether or
not the field theory can reproduce the genus $g$ anomaly equations of the B
model.

\vskip 1 cm

{\large \bf \noindent Acknowledgments}

We are grateful to Ori Ganor, Yaron Oz and Cobi Sonnenschein for many useful
discussions.

\newpage
{
\small
\parskip=0pt plus 2pt
\def\baselinestretch{1.}

}


\begin{thebibliography}{99}


\bibitem{B}
        E.~Witten,
        {\em ``Mirror manifolds and topological field theory'',\/}
        in {\em Essays on Mirror Manifolds,\/}
        ed. S.T.~Yau  (International Press, 1992).

\bibitem{A}
        E.~Witten,
        \cmp118,88,411.

\bibitem{kapl}
        L.J.~Dixon, V.~Kaplunovsky and J.~Louis,
        \np355,91,649.

\bibitem{F1}
        M.~Bershadsky, S.~Cecotti, H.~Ooguri and C.~Vafa,
        hep-th/9302103, \np405,93,279.

\bibitem{CS}
        E.~Witten,
        {\em ``\CS{} Gauge Theory as a String Theory'',\/}
        IAS preprint IASSNS--HEP--92--45, hep-th/9207094.

\bibitem{OS}
        E.~Witten,
        \np268,86,253.

\bibitem{KS}
        M.~Bershadsky, S.~Cecotti, H.~Ooguri and C.~Vafa,
        {\em ``\KS{} theory of gravity and exact results for quantum string
        amplitudes'',\/}
        Harvard preprint HUTP--93--A025, hep-th/9309140.

\bibitem{\KS}
        K.~Kodaira and D.C.~Spencer,
        \annmath67,58,328, \noj71,60,43;
        K.~Kodaira, L.~Nirenberg and D.C.~Spencer,
        \annmath68,58,450.

\bibitem{22}
        B.~Zumino,
        \pl87,79,203.

\bibitem{boundfer}
        J.~Isida and A.~Hosoya,
        \ptp62,79,544;\\
        A.~Barducci, R.~Casalbuoni and L.~Lusanna,
        \np180,81,141;\\
        N.~Marcus and A.~Sagnotti,
        \pl188,87,58;\\
        N.~Marcus, \ju{Fix all these}
        {\em ``Open strings and boundary fermions'',\/}
        in {\em ``String Theory'',\/}
        ed. C.~Procesi and A.~Sagnotti (Academic Press 1990),
        {\em ``Open strings and superstring sigma models with boundary
        fermions'',\/}
        Washington preprint 40423--10 P8.

\bibitem{index}
        S.~Cecotti, P.~Fendley, K.~Intriligator and C.~Vafa,
        hep-th/9204102, \np386,92,405.

\bibitem{ising}
        S.~Cecotti and C.~Vafa,
        hep-th/9209085, \cmp157,93,139.

\bibitem{topgrav}
        D.~Montano and J.~Sonnenschein,
        \np313,89,258;\\
        J.M.F.~Labastida, M.~Pernici and E.~Witten,
        \np310,88,611.

\bibitem{DVV}
        R.~Dijkgraaf, E.~Verlinde and H.~Verlinde,
        {\em ``Notes on Topological String Theory and 2--D Quantum
        Gravity'',\/} in
        {\em  String theory and quantum gravity,\/} ed.  M.~Green, R.~Iengo,
        S.~Randjbar-Daemi, E.~Sezgin and H.~Verlinde (World Scientific, 1991).

\bibitem{to appear}
        N.~Marcus,
        to appear.

\bibitem{lamba}
        J.M.F.~Labastida and P.M.~Llatas,
        \np379,92,220.

\bibitem{polybook}
        See, for example,
        A.M. Polyakov,
        {\em ``Gauge fields and strings''\/}
        (Harwood Academic, 1987).

\bibitem{MM}
        S.D.~Mathur and S.~Mukhi,
        \np302,88,130.

\bibitem{nicepolch}
        See, for example,
        J.~Polchinski,
        \cmp104,86,37.

\bibitem{witanom}
        E.~Witten,
        {\em ``Global anomalies in string theory'',\/}
        in {\em Symposium on Anomalies, Geometry and Topology,\/}
        ed. W.A.~Bardeen and A.R.~White (World Scientific, 1985).

\bibitem{AG}
        L.~\'Alvarez-Gaum\'e,
        \cmp90,83,161.

\bibitem{israelis}
        S.~Elitzur, Y.~Frishman, E.~Rabinovici and A.~Schwimmer,
        \np273,86,93.

\bibitem{somebook}
        R.C.~Gunning,
        {\em ``Lectures on modular forms''\/}
        (Princeton University Press, 1962).

\bibitem{\RS}
        D.B.~Ray and I.M.~Singer,
        \annmath98,73,154.

\bibitem{lamba2}
        J.M.F.~Labastida and M.~Mari\~no,
        {\em ``Type B topological matter, \KS{} theory, and mirror
        symmetry'',\/}
        Universidade de Santiago preprint IUS--FT/7--94, hep-th/9405151.

\bibitem{Gilkey}
        See Lemma 1.7.7 in {\em ``Invariance theory, the heat equation, and
        the Atiyah--Singer index theorem'',\/} P.B.~Gilkey (Publish or
        Perish Inc, 1984).

\bibitem{BF}
        J.M.~Bismut and D.S.~Freed,
        \cmp106,86,159, \ibid{} \noj107,86,103;\\
        J.M.~Bismut, H.~Gillet and C.~Soule,
        \ibid{} \noj115,88,49, 79, 301;\\
        J.M. Bismut and K.~K\"ohler,
        \jou{J.\ Alg.\ Geom}1,92,647.

\bibitem{Paul}
        D.~Friedan and P.~Windey,
        \np235,84,395.

\bibitem{Barton}
        See, for example, B.~Zwiebach,
        hep-th/9206084, \np390,93,33.

\bibitem{N=2} H.~Ooguri and C.~Vafa,
        \mpl5,90,1389, \np361,91,469, \ibid{} \noj367,91,83;\\
        N.~Marcus, \np387,92,263.

\bibitem{italians}
        M.~Bonini, E.~Gava and R.~Iengo,
        \mpl6,91,795.

\end{thebibliography}
\end{document}